\documentstyle[prl,aps,epsfig]{revtex}

\begin{document}

\twocolumn[\hsize\textwidth\columnwidth\hsize\csname @twocolumnfalse\endcsname

\title{Collision of 1.4 $M_{\odot}$ Neutron Stars: Dynamical or Quasi-Equilibrium?}

\author{Mark Miller${}^{(1)}$, Wai-Mo Suen${}^{(1,2)}$ and Malcolm
Tobias${}^{(1)}$}

\address{${}^{(1)}$McDonnell Center for the Space Sciences,
Department of Physics,
Washington University, St. Louis, Missouri 63130}

\address{${}^{(2)}$Physics Department,
Chinese University of Hong Kong,
Hong Kong}

\date{\today}
\maketitle

\begin{abstract}

Shapiro put forth a conjecture stating that neutron stars in head-on
collisions (infalling from infinity) will not collapse to black holes
before neutrino cooling, independent of the mass of the neutron stars.
In a previous paper we carried out a numerical simulation showing a
counter example based on 1.4 $M_{\odot}$ neutron stars, and provided an
analysis explaining why Shapiro's argument was not applicable for this
case.

A recent paper by Shapiro put forth an argument suggesting that
numerical simulations of the 1.4 $M_{\odot}$ collisions could not
disprove the conjecture with the accuracy that is presently
attainable.  We show in this paper that this argument is not
applicable for the same reason that the Shapiro conjecture is not
applicable to the 1.4 $M_{\odot}$ neutron star collision, namely, the
collision is too dynamical to be treated by quasi-equilibrium
arguments.

\end{abstract}


\narrowtext

\vskip2pc]

Shapiro in \cite{Shapiro98a} proposed an intriguing conjecture
on head-on collisions of neutron stars (NSs) infalling from
infinity.  It goes as follows: For two stable NSs that are described
by a polytropic equation of state (EOS) $P = K \rho ^ \Gamma$
[$K=K(s)$ is a function of the entropy $s$, the polytropic index
$\Gamma$ is a constant in space and time], it is conjectured that no
prompt collapse can occur for an arbitrary $\Gamma$ and an arbitrary
initial $K$, independent of the mass of the neutron stars.  The
basic argument of \cite{Shapiro98a} is that, with the polytropic EOS,
there always exists a stable equilibrium TOV configuration with the
same total mass and total energy as the two initially isolated TOV
stars.  As the mass and energy are conserved to a good extent in the
head-on collision process, Ref. \cite{Shapiro98a} concluded that the
merged object will be stable until energy is radiated away at a much
longer timescale.

In \cite{Miller99a}, we present a counter example to the
conjecture.  We carried out a numerical simulation of the head-on
collision of two 1.4 $M_{\odot}$ neutron stars, with polytropic
index of $2$ and initial polytropic coefficient $K=1.16 \times 10^5 \;
\frac {{cm}^5}{g \; s^2}$, a typical choice used in neutron star collision
studies (see e.g., \cite{Ruffert98a} and references therein).
We found that a black hole is formed promptly.  We also proposed a
reason why Shapiro's argument is not applicable to the 1.4 $M_{\odot}$
case (see below).

Recently Shapiro \cite{Shapiro99a} put forth an argument suggesting
that the $1.4 M_{\odot}$ simulation in \cite{Miller99a} could not
possibly have the accuracy needed to determine whether a black hole is
formed or not.  The argument is again based on stable equilibrium TOV
configurations.  It is pointed out that in order to determine whether
a TOV configuration is on the stable or unstable side of the
equilibrium curve near the critical point of stability, one has to
determine its total energy to very high accuracy.  For the 1.4
$M_{\odot}$ simulation reported in \cite{Miller99a}, it is estimated
that the energy must be calculated to within an accuracy of $0.5\%$
throughout the evolution.  It is claimed that such an accuracy is
beyond what one can achieve with a 3D code with presently available
computing power.

In the following we discuss why the observation in \cite{Shapiro99a},
although interesting and potentially important to other numerical
studies, is {\it irrelevant} to the simulation reported in \cite{Miller99a}.
In fact, the reason why this argument is irrelevant is the same as why
the Shapiro's argument is not applicable to the 1.4 $M_{\odot}$ case
in the first place, as given already in \cite{Miller99a}.  

In \cite{Miller99a} we pointed out that there is a hidden assumption
in Shapiro's argument, namely, that the collision
process can be approximated by a quasi-equilibrium process, in two
related senses: (A) The coalescing matter can be described by one {\it
single} EOS everywhere [$K(s)$ is a function of time but not space,
that is, $s$ is a spatial constant, uniform throughout the object],
and further, (B) whether it collapses or not is determined by the
hydro-static equilibrium condition, i.e., whether a stable equilibrium
TOV configuration {\it exists} or not.  We pointed out in
\cite{Miller99a} that this quasi-equilibrium assumption is not
self-evident for the head-on collision of the 1.4 $M_{\odot}$ NSs
after estimating the various time scales involved.  In the following
we expand on this question of ``dynamical'' vs. ``quasi-equilibrium''.

For concreteness and for the purpose of establishing one counter
example to Shapiro's conjecture, in both \cite{Miller99a} and the
present paper we focus on the 1.4 $M_{\odot}$ case, with a specific
EOS.  We shall not comment, nor make any claim, on any other
preliminary simulations or results with a different mass or EOS.

In \cite{Miller99a} we found that the quasi-equilibrium assumption
was broken in the sense of both (A) and (B) for the 1.4 $M_{\odot}$
case .  (A) is violated as
the coalesced object does not have time to thermalize
before it collapses.  Notice that with a strong shock,
``thermalization'' proceeds much faster than the heat conduction time
scale.  However the shock front is still not fast enough in its
outward propagation; it gets trapped inside the apparent horizon
before it can reach a good part of the infalling matter.  The
polytropic coefficient $K(s)$ is never a spatial constant, nor approximately
constant, throughout the coalesced object.  This implies that (B) must
also be violated: the collision process is so dynamical that although
a stable equilibrium state exists (allowed by the energy and mass
conservation laws), it is {\it not attained} in the collapse process.

One central message of \cite{Miller99a} and the present paper is that
the head-on collision of 1.4 $M_{\odot}$ NSs is so dynamical that one
must not use any notion of quasi-equilibrium,
on which both the original Shapiro conjecture and the ``accuracy
argument'' are based.  

We have already provided evidence for this dynamical nature of the
collision in \cite{Miller99a}.  To strengthen this message of
``dynamical'' vs. ``quasi-equilibrium'', we focus on one specific aspect
in this paper.  When the two 1.4 $M_{\odot}$ NSs touch after infalling
from infinity, they approach one another at a fraction of the light
speed.  A strong shock is generated, converting about $90\%$ of the
bulk kinetic energy to thermal energy in the post-shock material.
However, the rest of the stars (pre-shock) cramp in so rapidly that an
apparent horizon quickly forms, trapping everything, including the
shock front, inside.  In Fig. 1 below, the solid line is the world
line of the shock front in the $z$ direction (direction of infall),
plotted in coordinate distance and coordinate time.  The ``*'' 
represents the location of the apparent horizon found in the
simulation.  The dotted line represents one leg of the backward light
cone, starting backward from the apparent horizon.  The backward light
cone meets the shock front at $t=0.195 ms$, which is $0.048 ms$ after
the two stars first touched.  At this point the shock has just barely
propagated outward and the material heated up by the shock wave (the
post-shock material) makes up only a small fraction of the total
amount of matter on this time slice.  In terms of the number of
baryons, it is less than 5\%.   
(One could consider different time slices, but
the basic picture is the same.)  We do not expect that such an
insignificant fraction of heated material could have much effect on
the overall dynamics of the infalling matter, let alone providing the
thermal pressure to prevent the collapse as envisioned in the Shapiro
conjecture.  At later times the shock heated material does make up a
bigger fraction of the total mass.  However, from the point of
intersection of the dashed and solid line at $t=0.195 ms$ onward, the
shock heated material is {\it causally disconnected} with the apparent
horizon shown and could not affect its formation.  This highlights how
far away from thermal and dynamical equilibrium the collision process
is.

\begin{figure}
\psfig{figure=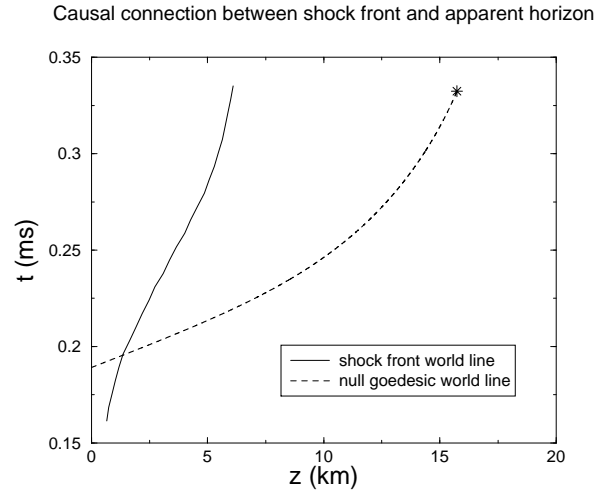,width=8.0cm}
\caption{ The causal relation between the apparent horizon and the
   shock front is displayed.  The point marked by ``*'' represents the
   intersection of the apparent horizon with the $+z$ axis (the stars
   are falling towards one another along the $z$ axis with the center
   of collision at $z=0$).  A null geodesic traveling along the $z$
   axis is shown as a dotted line.  The solid line represents the
   world line of the intersection of the shock front on the $+z$
   axis.  World lines are plotted in coordinate distance
   vs. coordinate time.}  
\end{figure}

It may be worthwhile to point out that the features of this figure
have been quite carefully examined.  The position of the AH has been
subjected to convergence and stability tests with respect to
different boundary conditions and different locations of the outer
boundary.  The existence of trapped surfaces in the spacetime have
been explicitly verified.  To confirm the shock propagation speed in
the curved spacetime simulation, we have carried out tests in which we
extracted the proper density, pressure and velocity of the fluid flow
on two sides of the shock at various times in the collision
simulation, and set up shock tube tests with the same hydrodynamic
conditions in flat space.  Such tests are meaningful because of the
fact that shock propagation is a local phenomena, and that the flat
space shock treatments in our code has been thoroughly tested
previously \cite{Font98b}.  The details of these tests will be given in
a follow up paper.

The main aim of this paper is to contrast the dynamical picture
demonstrated in our simulation with the ``equilibrium picture'' of
\cite{Shapiro98a,Shapiro99a}.  In the ``equilibrium picture'' one
envisions that the shock wave bounces a couple of times across the whole
coalesced object, heating it up approximately uniformly [so that the
polytropic constant $K(s)$ becomes a spatial constant], and the
coalesced object can be described by a TOV configuration in
equilibrium.  Refs. \cite{Shapiro98a} and \cite{Shapiro99a} then
analyzed the stability of this ``resulting'' TOV configuration and
concluded that it could not collapse.

The argument of \cite{Shapiro98a} and \cite{Shapiro99a} would be strictly
valid if the collision process is in quasi-equilibrium instead of freefall.
In Fig. 2, panel (A1) represents two NSs in free fall towards
each other.  Panel (A2)
represents the shock wave heating up the merging stars (with the dark grey
area representing the material at higher temperature and hence higher
$K$).  However, at this point the apparent horizon pops into existence
engulfing most of the infalling matter.  
(A3) represents the black hole finally formed.  Fig. 3 represents a 
thought experiment in which the quasi-equilibrium analysis is valid.
Panel (B1) represents two NSs with ropes attached slowly
lowered towards one another, with the potential energy extracted
and re-deposited back into the stars.  This makes the stars gradually
heat up, maintaining quasi-equilibrium throughout the process.
(To be precise, each fluid element has to be tied and arranged into
its quasi-equilibrium position at each point in time.)  This is
represented in (B2).  Darker grey is used to represent the uniformly heated
object with the potential energy extracted and redistributed back
to it through the ropes.  One final equilibrium (hot) TOV star is formed,
represented by (B3).  The stability of this TOV star
is determined by the considerations in
Refs. \cite{Shapiro98a} and \cite{Shapiro99a}.

\begin{figure}
\psfig{figure=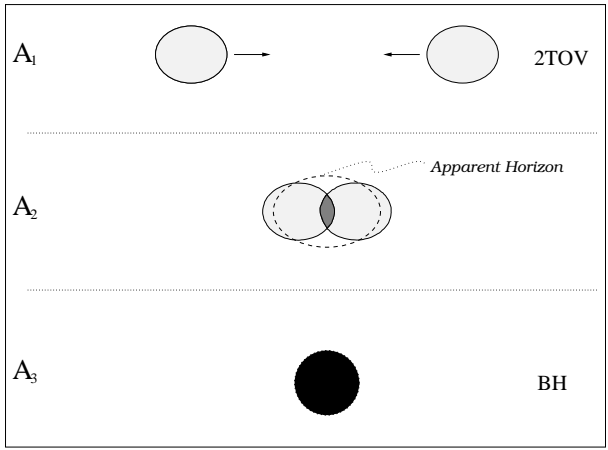,width=8.0cm}

\vspace{0.2cm}

\caption{The freefall collision of two 1.4
$M_{\odot}$ NSs.  In (A1) two NSs are in free fall towards each
other. The grey region in panel (A2) represents the material heated by
the shock wave.  At this point, an apparent horizon forms, engulfing
most of the infalling matter.  Panel (A3) represents the black hole
finally formed.  }
\end{figure}

\begin{figure}

\psfig{figure=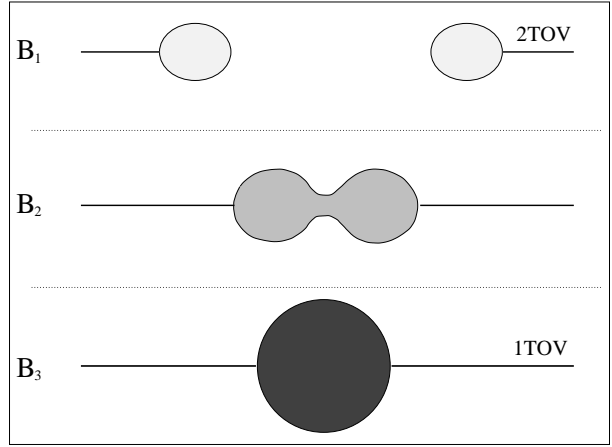,width=8.0cm}

\vspace{0.2cm}

\caption{The merging of two NSs in
quasi-equilibrium.  (B1) represents two NSs with ropes attached
slowly lowered towards one another, with the potential energy
extracted and re-deposited back into the stars.  (The fluid elements are
tied and arranged into their quasi-equilibrium position throughout the
process.)  Panel (B2) represents the uniformly heated, merging
object with the potential energy extracted and redistributed back to it through
the ropes.  Panel (B3) represents the final equilibrium TOV star
formed.  In this quasi-equilibrium scenario the considerations in
Refs. {\protect \cite{Shapiro98a}} and {\protect \cite{Shapiro99a}} 
would be strictly applicable.}
\end{figure}

Fig. 4 schematically represents the configuration space of the coupled
Einstein-GRHydro equations.  The lower left dot denoted 2TOV
represents the state of two TOV configurations infinitely
separated.  In free fall governed by the coupled Einstein-GRHydro
equations, it evolves along the solid line towards the lower right
hand dot denoted BH representing the solution of a black hole.  The
three stages A1, A2 and A3 in Fig. 2 are labeled.  The dotted line
represents the evolution of the Panels B in Fig. 3 (the
Einstein-GRHydro equations with source terms including the ropes): The
two TOV stars are tied and lowered towards one another.  This leads to
the state represented by the dot in upper right hand corner denoted
1TOV.  The argument in Ref. \cite{Shapiro99a} amounts to pointing out
that should one actually carry out a numerical simulation following
this dotted line, one would have to maintain $0.5\%$ accuracy in order
to determine whether the 1TOV solution obtained at B3 is stable or
not.  We see that this accuracy requirement is irrelevant to the
head-on collision study: We are not carrying out a simulation along
the quasi-equilibrium path (if such a simulation is at all
possible).

\begin{figure}
\psfig{figure=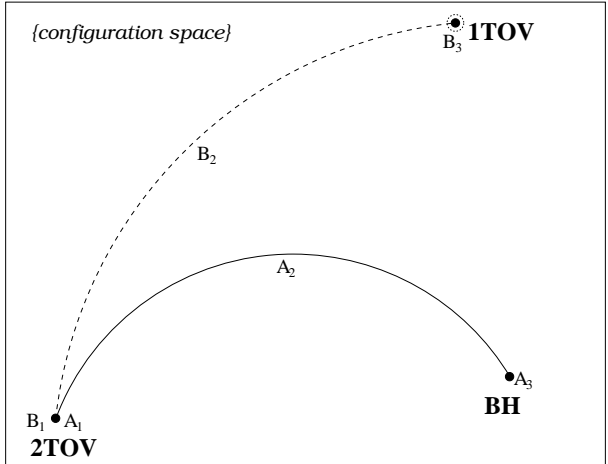,width=8.0cm}

\caption{This figure represents the configuration space of the coupled
Einstein-GRHydro equations.  The lower left dot denoted 2TOV
represents the state of two TOV configurations infinitely
separated.  In free fall governed by the coupled Einstein-GRHydro
equations, it evolves along the solid line towards the lower right
hand dot denoted BH representing a black hole.  The three stages A1,
A2 and A3 in Fig. 2 are labeled.  The dotted line represents the
evolution in quasi-equilibrium depicted in Panels B of Fig. 3: The
two TOV stars are tied and lowered towards one another.  This leads to
the state represented by the dot in upper right hand corner denoted
1TOV.
}
\end{figure}

How accurate do we have to get in following the evolution depicted by
the solid line in Fig. 4 before we can be sure that a black hole is
actually formed?  We do not have an answer to this question presently.
The accuracy we achieved can be monitored by the Hamiltonian
constraint violation at the point we found the AH (along the z-axis).
The Hamiltonian constraint violation is $0.146, 0.096, 0.062, 0.040$
(in units of $16 \pi \rho$, with $\rho$ being the maximum
rest mass density at that time), 
for the simulations carried out with grid sizes of $40^3,
96^3, 160^3, 256^3$, respectively.  The simulations are carried
out with only one octant of the grid evolved.  (Due to the symmetry of the
problem, only one octant of the domain needs to be numerically
calculated.  The $256^3$ simulation corresponds to having about 140
points across the diameter of one star.)  We note two points: 1. the
formation of a black hole can be obtained by a simulation with rather
coarse resolution, and 2. this feature of black hole formation is
stable with respect to increasing resolution, i.e., a long time scale
convergence test.  We emphasize that in numerical studies one {\it must}
insist that the physical feature one is studying is subjected
to, and passes convergence tests.  We emphasize that here we
are talking about {\it not} just the usual short time convergence code
test.  The convergence tests must be carried out for the specific system of
physical interest, and {\it maintained} throughout the time of
evolution up to the time the physical feature under consideration is
extracted: a ``long term'' convergence test.  Together with the
consistency tests making sure that the finite difference equations are
faithful to the differential equations, one can then invoke the Lax
theorem to give oneself a reasonable confidence on the numerical
result.  This is a point that cannot be over-emphasized for all
numerical studies.  All simulations discussed in this paper and in
\cite{Miller99a} have gone through these consistency and long term
convergence tests.

We further note that even though the simulations we carried out may be
reliable and show the prompt formation of a black hole, they still may
not constitute counter examples to the Shapiro conjecture, subject to
the following consideration.  In our numerical simulations, the
initial data is set with the NSs at a finite separation with an infall
velocity.  The initial data set we used may not be the same as what it
would actually be falling in from infinity.  Indeed, setting initial
data in numerical simulations to represent a given physical scenario
is a major problem in numerical relativity.  Fortunately, the problem
at hand is considerably easier than trying to set up initial data to,
say, obtain a gravitational waveform to compare with observations.
There, one has to determine precisely the correct initial data
corresponding to the physical scenario.  Here, to establish a counter
example to the Shapiro conjecture, our strategy is to construct 
reasonable initial data sets that one can consider as {\it
approximately} representing the 1.4 $M_{\odot}$ head-on collision
problem.  If {\it all} of them leads to the same dynamical picture of
evolution with the same qualitative behavior, and all of them produce
a black hole promptly, we are willing to conclude that a counter
example is established, even though we might not be able to pin down
precisely the exact initial data.

For example, it is difficult if at all possible to determine the exact
infalling velocity at the point we start the evolution (typically the
initial separation is chosen to be around $45km$ in our numerical
study).  We choose the initial velocity to be given by the Newtonian
freefall velocity ($M/r$ at $45km$ separation is about $0.05$ for 1.4
$M_{\odot}$ and the EOS used; it is not too much out of the Newtonian
regime).  We have also carried out simulations with $10\%$ higher
and lower initial velocities, and confirmed that the prompt collapse
result is not sensitive to the exact choice.  Although we still cannot
pin down the ``correct'' velocity, we believe that this
inability to pin it down will not affect our conclusion of prompt
collapse.  

For this reason we have also carried out simulations using different
constructions of the initial metric and matter distributions.  There
are different ways to put two 1.4 $M_{\odot}$ TOV configurations
together.  One can directly put the TOV density profile in and
solve the four initial data constraints, or one can require the baryon
number of the data to be held fixed before and after the solving of the
initial data constraints, or the total ADM mass.  While holding the
total baryon number is arguably ``preferable'', we have also carried
out simulations of the other setups.  The details of the evolutions
with different initial data sets will be given in a follow up
paper; but in all cases studied, we found the same qualitative
behavior, and all of them produce a black hole promptly.  It is with
these tests we feel we have quite confidently established a counter
example to the Shapiro conjecture: Even though we cannot pin down
the exact initial data, the dynamical nature of the problem
and the prompt collapse to a black hole appear to be generic features and are
insensitive to details in choosing initial data.

Ref. \cite{Shapiro99a} also raised the point that we have not
considered tidal distortion of the two stars in the initial data, and
that such an error could invalidate the prompt collapse result in view of
the ``$0.5\%$ accuracy requirement''.  In the above we pointed out
that the ``$0.5\%$ accuracy requirement'' is irrelevant to the present
consideration, but it is true that we have not carried out studies of
the tidal distortion effect.  We note that we could have incorporated
tidal distortion with the same type of argument as above: Carry out
simulations with different distortions and verify that the qualitative
behavior of the collapse is the same.  Also, one can conceivably
construct a tidally distorted initial configuration using the
``conformally flat quasi-equilibrium treatment'' minimizing the energy
(for review, see e.g., Refs. \cite{Gourgoulhon98a,Rasio99}),
although quasi-equilibrium for the head-on case would be less accurate
compared to the inspiraling case at the same initial separation.  We
have not investigated along this line because of the following
consideration: The spherical symmetric density distribution of the
individual star we used in starting off the initial data calculation
has more energy compared to the ``correctly tidally distorted''
configuration that one would have obtained by tracking the stars all
the way in from infinity.  The spherical symmetric density
distribution we used is in fact a ``distorted configuration'' with
respect to the correct density distribution the stars would have
infalling from infinity.  By starting off at a finite distance with
spherical symmetric distributions we have put {\it more} potential
energy in the system than it would actually have.  Hence, according to
Shapiro's argument, this should lead to {\it more} thermal energy upon
coalescence and it should make it {\it more} difficult to
collapse, if it has any effect at all.  But still it promptly
collapses even in this case of more energy.  With this consideration
in mind, we felt that tidal distortion was not a major concern towards
the goal of establishing a counter example to the Shapiro conjecture.
(It would be an important concern if one were trying to determine the
gravitational waveform for the collision.)

\paragraph*{\bf Conclusion.}
\label{time}

In the two recent papers \cite{Shapiro98a} and \cite{Shapiro99a},
Shapiro provided useful insight to processes involving neutron stars.
However, the arguments in these papers are not applicable to the
head-on collision of 1.4 $M_{\odot}$ neutron stars in \cite{Miller99a}
and this paper.  The crux of the problem is that such a collision is
too dynamical to be studied using quasi-equilibrium
analyses.

\paragraph*{\bf Acknowledgments.}

This research is supported by NASA NCS5-153, NSF NRAC MCA93S025, and
NSF grants PHY96-00507, 96-00049, and 99-79985.

\bibliographystyle{prsty}


\end{document}